\documentstyle{article}
\title{\bf On the geometry of the Hawking Turok instanton }
\author{ Pedro F. Gonz\'alez-D\'{\i}az \footnotemark \\
Instituto de Matem ticas y F¡sica Fundamental,\\
Consejo Superior de Investigaciones Cient¡ficas, Serrano 121,
28006 Madrid, Spain
}
\date{March 20, 1998}
\begin{document}
\maketitle
\large
\setlength{\baselineskip}{0.5cm}

\footnotetext{
Email: iodpf21@cc.csic.es }

\begin{center}
{\bf Abstract}
\end{center}

We discuss the coordinate complexification necessary to
obtain a spatially homogeneous open universe in the
singular instantonic solution recently suggested by
Hawking and Turok, concluding that only the time
coordinate $\sigma$ of the spatially inhomogeneous
De Sitter-like Lorentzian solution and its scale
factor should be rotated
in order to continue into a Lorentzian open universe.

\vspace{1cm}

The need for formulating open inflationary models with
$\Omega<1$ comes mainly from recent estimates of the
present Hubble constant [1-4] and density of gravitational
lenses in the universe [5,6]. However, most of the known
inflationary models predict a value for $\Omega$ very
close to unity and cannot, therefore, conform these new
observations. It is a rather ironic fact that the
first proposed solution to the percolation problem of
the original Guth scenario for inflation was given in
terms of an open inflationary model [7], but this
paper was largely darkened by the almost inmediate
publication of the proposal for the new inflationary
model [8,9].
Quite recently a new approach to solve this problem
has been advanced by Hawking and Turok [10], based on the
idea that open homogeneous universes can be created from
nothing, without an intermediate stage involving false
vacuum, using the Hartle-Hawking no boundary [11] and a
rather generic potential. The appearance of this
seminal paper has already prompted an influx of
comments, criticisms and new work on the subject [12].

The way in which one can construct the geometry of the
Hawking-Turok instanton has also been discussed by
Unruh [13]. Here, we would like to contribute the latter
discussion, clarifying the meaning of some details on
the complexification
method employed in the construction of the instanton.
Such a method starts with the Euclidean solution to
Einstein equations [10]
\begin{equation}
ds^2=d\sigma^2+b(\sigma)^2\left(d\psi^2
+\sin^2\psi d\Omega_2^2\right),
\end{equation}
with $d\Omega_2^2$ the metric on the unit two-sphere.
In order to convert solution (1) into the metric for
an open spatially homogeneous universe with real scale
factor, Hawking and Turok used two separate
continuations [10].
First, they matched (1) to the spatially inhomogeneous
De Sitter-like Lorentzian solution
\begin{equation}
ds^2=d\sigma^2+b(\sigma)^2\left(-d\tau^2
+\cosh^2\tau d\Omega_2^2\right),
\end{equation}
accross the surface $\psi=\frac{\pi}{2}$ by making the
continuation $\psi=\frac{\pi}{2}+i\tau$, and then they
continued (2) into the metric of a Lorentzian isotropic
open universe,
\begin{equation}
ds^2=-dt^2+a(\sigma)^2\left(d\chi^2
+\sinh^2\chi d\Omega_2^2\right),
\end{equation}
by using the additional continuation
\[\tau=\frac{\pi}{2}i+\chi ,\;\; \sigma=it ,\;\;
b(it)=ia(t) \]
at the singular surface $\sigma=0$ where $b=\sigma$.

Since in the coordinates $\sigma$, $\tau$ of metric (2)
$\sigma=0$ is a null surface which is similar to the
horizon in Rindler coordinates or the Schwarzschild
solution, Unruh has pointed out [13] that the second of the
above two continuations is not necessary and that it should
be replaced for a coordinate extension similar to that
of Kruskal for black holes, so going beyond the singular
surface $\sigma=0$ into the region described by a metric
like (3). In order to implement such an extension he
introduced a conformal coordinate defined by
\[\Sigma=\int_0^\sigma\frac{d\sigma '}{b(\sigma ')}\]
and Kruskal-like coordinates $U=e^{-\tau+\Sigma}$,
$V=-e^{\tau+\Sigma}$. One can then obtain from (2)
\begin{equation}
ds^2=\frac{b(\sigma)^2}{e^{2\Sigma}}
\left(-dUdV+\frac{1}{4}(U-V)^2 d\Omega_2^2\right).
\end{equation}
Denoting $F=\frac{b(\sigma)^2}{e^{2\Sigma}}$ and introducing
new coordinates $T$ and $\chi$ through $U=e^{T-\chi}$,
$V=e^{T+\chi}$, one finally get for the metric
\begin{equation}
ds^2=Fe^{2T}\left(-dT^2+d\chi^2+\sinh^2\chi d\Omega_2^2\right).
\end{equation}
Then, Unruh claimed a metric like (5)
to be the equation for an open
spatially homogeneous universe with scale factor
$e^T\sqrt{F}$. However, it can be seen that the function
$F$ is given by $F=-b^{2}e^{-2T}$ and thereby is negative
definite, so that the scale
factor in (5) is pure imaginary and hence the extension
across the singularity at $\sigma=0$ used by Unruh
actually implies some complexification of the coordinates
involved at metric (5). One can readily see that, in order for
metric (5) to really be the equation for an open
spatially homogeneous universe with real, positive definite
scale factor and real coordinates, the conformal coordinate
$\Sigma$ and the Kruskal-like coordinates $U$, $V$ need to
be kept all real and unchanged, so that finally $T$ and
$\chi$ would remain real as well. Clearly, this can only
be achieved if we simultaneously rotate $\sigma$ and $b$
so that $\sigma=it$ and $b(it)=ia(t)$, such as Hawking and
Turok originally did [10], though contrary to these authors,
we must keep coordinate $\tau$ real and unchanged, instead
of comlexifying it with a constant imaginary term
$\frac{\pi}{2}i$ [10]. Thus,
although the extension suggested by Unruh relaxes the part of
the complexification that corresponds to this constant term,
it cannot avoid the rest of it.

In summary, what the solution of Hawking and Turok requires
to lead to an open universe in a natural way is, on the
one hand, the complexification from the real to the
Euclidean metric and, on the other hand, an extension
into the open universe sector which should be accompanied
by a rotation of the scale factor induced only by the
Wick rotation of the original $\sigma$ coordinate.

\vspace{.5cm}

\noindent {\bf Acknowledgements}
For useful correspondence the author thanks W. Unruh.
This research was supported by DGICYT under research
project N§ PB94-0107.

\noindent\section*{References}

\begin{description}

\item [1] A.G. Riess, W.H. Press and R.P. Kirshner, Ap. J.
438, L17 (1995).
\item [2] J. Mould and W. Freedman, Nature 381, 555 (1996)
\item [3] A. Sandage et al., Nature 381, 555 (1996).
\item [4] T. Kundic et al., Ap. J. 482, 75 (1997).
\item [5] E.L. Turner, Ap. J. 365, L43 (1990).
\item [6] M. Fukugita, T. Futamase and M. Kasai,
Mon. Not. R. Astron. Soc. 246, 24P (1990).
\item [7] J.R. Gott, Nature 295, 304 (1982); astro-ph/9712344
\item [8] A.D. Linde, Phys. Lett. B108, 389 (1982).‡
\item [9] A. Albrecht and P.J. Steinhardt, Phys. Rev. Lett.
48, 1220 (1982).
\item [10] S.W. Hawking and N. Turok, hep-th/9802030, to
appear in Phys. Lett. B.
\item [11] J.B. Hartle and S.W. Hawking, Phys. Rev. D28, 1960
(1983).
\item [12] A.D. Linde, gr-qc/9802038; S.W. Hawking and N. Turok,
gr-qc/9802062; A. Vilenkin, hep-th/9803084; W.Z. Chao,
hep-th/9803121; R. Bousso and A.D. Linde, gr-qc/9803068;
N. Turok and S.W. Hawking, hep-th/9803156 .
\item [13] W. Unruh, gr-qc/9803050.
\end{description}

\end{document}